\documentclass[12pt,aps,onecolumn,superscriptaddress]{revtex4}

\usepackage{graphicx}
\usepackage{amsmath}
\usepackage{hyperref}

\newcommand{\widebar}[1]{\ensuremath{\overset{\hspace{0.8mm}\underline{\hspace{2.33mm}}}{#1}}}
\newcommand{\var}{} % auxiliary variable

\begin{document}

\title{\large\bf\boldmath
Measurement of the $e^+e^- \to \pi^+\pi^-\pi^0\eta$ cross section below $\sqrt{s}=2$~GeV}

%==============================================
\newcommand{\binp}{\affiliation{Budker Institute of Nuclear Physics, SB RAS, Novosibirsk 630090, Russia}}
\newcommand{\nsu}{\affiliation{Novosibirsk State University, Novosibirsk 630090, Russia}}
\newcommand{\nstu}{\affiliation{Novosibirsk State Technical University, Novosibirsk 630092, Russia}}

\author{M.~N.~Achasov} \binp\nsu
\author{A.~Yu.~Barnyakov} \binp\nsu\nstu
\author{K.~I.~Beloborodov} \binp\nsu
\author{A.~V.~Berdyugin} \binp\nsu
\author{A.~G.~Bogdanchikov} \binp
\author{A.~A.~Botov} \email[e-mail: ]{A.A.Botov@inp.nsk.su}\binp
\author{T.~V.~Dimova} \binp\nsu
\author{V.~P.~Druzhinin} \binp\nsu
\author{V.~B.~Golubev} \binp\nsu
\author{L.~V.~Kardapoltsev} \binp\nsu
\author{A.~G.~Kharlamov} \binp\nsu
\author{A.~A.~Korol} \binp\nsu
\author{S.~V.~Koshuba} \binp
\author{D.~P.~Kovrizhin} \binp
\author{A.~S.~Kupich} \binp\nsu
\author{R.~A.~Litvinov} \binp
\author{A.~P.~Lysenko} \binp
\author{K.~A.~Martin} \binp
\author{N.~A.~Melnikova} \binp\nsu
\author{N.~Yu.~Muchnoy} \binp\nsu
\author{A.~E.~Obrazovsky} \binp
\author{E.~V.~Pakhtusova} \binp
\author{E.~A.~Perevedentsev} \binp\nsu
\author{K.~V.~Pugachev} \binp\nsu
\author{S.~I.~Serednyakov} \binp\nsu
\author{P.~Yu.~Shatunov} \binp
\author{Yu.~M.~Shatunov} \binp\nsu
\author{D.~A.~Shtol} \binp
\author{Z.~K.~Silagadze} \binp\nsu
\author{A.~N.~Skrinsky} \binp
\author{I.~K.~Surin} \binp
\author{Yu.~A.~Tikhonov} \binp\nsu
\author{A.~V.~Vasiljev} \binp\nsu
\author{I.~M.~Zemlyansky} \binp

\begin{abstract}
The process $e^+e^- \to \pi^+\pi^-\pi^0\eta$ is studied in the center-of-mass energy region
below 2~GeV with the SND detector at the VEPP-2000 $e^+e^-$ collider.
The four intermediate states contribute to this process: $\omega\eta$, $\phi\eta$,
$a_0(980)\rho$, and a structureless $\pi^+\pi^-\pi^0\eta$ state.
We measure the total $e^+e^- \to \pi^+\pi^-\pi^0\eta$ cross section and the cross sections for
its components: $\omega\eta$, $\phi\eta$, and a sum of $a_0(980)\rho$ and the structureless
state.
Our results are in agreement with previous measurements and have comparable or better
accuracies.
\end{abstract}

\maketitle

%==============================================
\section{Introduction}

The main goal of experiments at the VEPP-2000 $e^+e^-$ collider~\cite{VEPP} is the measurement
of the total cross section of $e^+e^-$ annihilation into hadrons.
The information on the cross section is necessary, in particular, for calculation of the
hadronic vacuum polarization contribution into the muon anomalous magnetic moment and the
running electromagnetic coupling constant.
Below 2~GeV the total hadronic cross section is determined as a sum of exclusive cross sections
for all possible hadronic modes.
At the moment most of these cross sections have been measured.
At the same time still there are processes, which cross sections have not been measured yet or
have been measured with insufficient accuracy: $e^+e^- \to \pi^+\pi^-\pi^0\eta$,
$\pi^+\pi^-3\pi^0$, $\pi^+\pi^-2\pi^0\eta$, $\pi^+\pi^-4\pi^0$, etc.

In this work we analyze the process
\begin{equation}
  e^+e^- \to \pi^+\pi^-\pi^0\eta
  \label{EQ:ETA3PI}
\end{equation}
using a data sample collected with the Spherical Neutral Detector (SND)~\cite{SND} at the
VEPP-2000 $e^+e^-$ collider in the center-of-mass (c.m.) energy region $E = \sqrt{s} < 2$~GeV.
The intermediate states contributing to this process were established in the SND preliminary
analysis~\cite{MESON2016} and in the CMD-3 work~\cite{CMD}.
They are the following:
\begin{eqnarray}
  e^+e^- & \to & \omega\eta \to 3\pi\eta, \label{EQ:OMETA}\\
  e^+e^- & \to & \phi\eta \to 3\pi\eta, \label{EQ:PHIETA}\\
  e^+e^- & \to & a_0(980)\rho \to 3\pi\eta, \label{EQ:ARHO}\\
  e^+e^- & \to & nres \to 3\pi\eta. \label{EQ:NRES}
\end{eqnarray}
The latter process~(\ref{EQ:NRES}), marked as $nres$, does not reveal any clear structure, and
may proceed via $\rho(1450)\pi$ intermediate state, which is difficult to be identified.
The $e^+e^- \to \omega\eta$ cross section was previously measured in the BABAR~\cite{OMETA},
SND~\cite{OMETA2}, and CMD-3~\cite{CMD} experiments.
The $e^+e^- \to \phi\eta$ process was studied in the BABAR experiment in several $\phi$ and
$\eta$ decay modes~\cite{PHIETA,PHIETA2,PHIETA3}, and in the SND experiment in the
$K^+K^-\gamma\gamma$ final state~\cite{PHIETA4}.
The study of the $e^+e^- \to \pi^+\pi^-\pi^0\eta$ process was performed by the CMD-3
detector~\cite{CMD}, which measured for the first time the total $e^+e^- \to
\pi^+\pi^-\pi^0\eta$ cross section in the energy region $E < 2$~GeV and the cross sections for
the subprocesses~(\ref{EQ:ARHO}) and~(\ref{EQ:NRES}).

%==============================================
\section{Data and simulation}
\label{DATA}

In this work we analyze a data sample with an integrated luminosity of 27~pb$^{-1}$ recorded by
the SND detector~\cite{SND} at the VEPP-2000 $e^+e^-$ collider in 2011 and 2012.
In the energy range under study, 1.34--2.00~GeV, data were collected in 36 energy points.
The c.m.\ energies for these points were determined with accuracy of 2--6~MeV by the CMD-3
detector, which collected data simultaneously with SND, using momentum measurements in Bhabha
and $e^+e^- \to p\bar{p}$ events~\cite{BEAM}.
Because of the absence of any narrow structures in the cross sections under study, the 36
energy points are merged into 13 energy intervals listed in Table~\ref{TAB:EXP}.
The weighted average energy for each interval also listed in the Table~\ref{TAB:EXP} is
calculated taking into account the luminosity distribution over the energy points entering into
the interval.

Simulation of the processes $e^+e^- \to \omega\eta$ and $\phi\eta$ is done with Monte Carlo (MC)
event generators based on the $e^+e^- \to VP$ model with $V \to \rho\pi \to \pi^+\pi^-\pi^0$,
where $V$ is a vector meson and $P$ is the $\eta$-meson.
In simulation of the $a_0(980)\rho$ intermediate state, it is assumed that the $a_0(980)$ and
$\rho$ are produced in the $S$-wave.
The simulation of the structureless process $e^+e^- \to nres \to 3\pi\eta$ is performed in the
hypothesis $e^+e^- \to \rho(1450)\pi$ with $\rho(1450) \to \rho\eta$.
The simulation of the main background process $e^+e^- \to \pi^+\pi^-\pi^0\pi^0$ is performed
according to Ref.~\cite{4PI}.
The event generators take into account radiative corrections to the initial particles
calculated according to Ref.~\cite{KURAEV}.
The angular distribution of additional photons radiated by the initial electron and positron is
simulated according to Ref.~\cite{BONNEAU}.
The energy dependencies of Born cross sections needed for calculations of the radiative
corrections are obtained from data using an iterative procedure.
As a first approximation we take the cross sections from Ref.~\cite{OMETA2} for the
$\omega\eta$, from Ref.~\cite{PHIETA} for the $\phi\eta$, and from Ref.~\cite{CMD} for the
$a_0\rho$ and $nres$ intermediate states.
In this work we also measure the total $e^+e^- \to \pi^+\pi^-\pi^0\eta$ cross section and the
cross section for the sum of the $a_0\rho$ and $nres$ intermediate states.
To determine the energy dependence of the detection efficiency for these processes, simulation
in the $e^+e^- \to \rho(1450)\pi$, $\rho(1450) \to \rho\eta$ model is used.
The initial Born cross sections are taken from Refs.~\cite{MESON2016,CMD}.
The iterative procedure is following.
From the MC simulation we determinate the detection efficiency and measure the Born cross
section from our data as described in Sec.~\ref{CRS}.
The obtained cross section is used in the second-iteration simulation.
We do not perform full simulation, but instead reweight simulated events using the new Born
cross section.
The iterations are stopped when the relative difference in the efficiency between two
successive iterations is less than 1\%.

Interactions of the generated particles with the detector materials are simulated using GEANT4
software~\cite{GEANT4}.
The simulation takes into account variation of experimental conditions during data taking, in
particular dead detector channels and beam-induced background.
The beam background leads to the appearance of spurious photons and charged particles in
detected events.
To take this effect into account, simulation uses special background events recorded during
data taking with a random trigger, which are superimposed on simulated events.

The integrated luminosity is measured using Bhabha scattering events with a 1\% systematic
uncertainty~\cite{OMETA2}.
The luminosity for the 13 energy intervals is listed in Table~\ref{TAB:EXP}.

%==============================================
\section{Event selection}
\label{SELECTION}

The process $e^+e^- \to \pi^+\pi^-\pi^0\eta$ is analyzed in the decay mode $\eta \to \gamma\gamma$.
Therefore, we select events with
\begin{itemize}
  \linespread{1}\selectfont
  \item two or three charged tracks originated from the interaction region;
  \item at least four photons with energy greater than 20~MeV;
  \item the energy deposition in the calorimeter greater than 300~MeV.
  \label{EN:SELPRELIM}
\end{itemize}
For selected events the vertex fit is performed using the parameters of two charged tracks.
The quality of the fit is characterized by the parameter $\chi^2_r$.
If there are three charged tracks in an event, the two of them with the lowest $\chi^2_r$ value
are selected.
The found vertex is used to refine the measured angles of charged particles and photons.
Then we select events containing at least one $\pi^0$ candidate and one $\eta$ candidate,
which are defined as two-photon pairs with invariant masses in the windows $70 < m_{12} <
200$~MeV/c$^2$ and $400 < m_{34} < 700$~MeV/c$^2$, respectively.
For these events, a kinematic fit to the $\pi^+\pi^-\pi^0\gamma\gamma$ hypothesis is performed
with the four constraints of energy and momentum balance, and the $\pi^0$ mass constrain for
the $\pi^0$ candidate.
The $\chi^2$ of the fit ($\chi^2_{3\pi2\gamma}$) is required to be less than 30.
For events containing more than one combination of $\pi^0$ and $\eta$ candidates, the
combination with the smallest value of $\chi^2_{3\pi2\gamma}$ is chosen.
The photon parameters after the kinematic fit are used to recalculate the $\eta$-candidate
invariant mass ($M_\eta$).
It is required to be in the window 400--700~MeV/c$^2$.
The event is then refitted with the additional $\eta$-mass constraint.
The refined energy of the $\eta$-meson candidate is used to calculate the invariant mass of the
system recoiling against the $\eta$ meson ($M_\eta^{rec}$).

The main background source for the process under study is the process
\begin{equation}
  e^+e^- \to \pi^+\pi^-\pi^0\pi^0.
  \label{EQ:4PI}
\end{equation}
For its suppression, a kinematic fit to the $e^+e^- \to \pi^+\pi^-\pi^0\pi^0(\gamma)$
hypothesis is performed, in which radiation of an additional photon along the beam axis is
allowed.
The $\pi^0$ candidates are defined as two-photon pairs with invariant masses in the window
70--200~MeV/c$^2$.
The fit requires the total energy and momentum balance and two $\pi^0$-mass constraints.
The energy of the additional photon moving along the beam axis is determined from the fit.
The condition $\chi^2_{4\pi(\gamma)} > 200$ is applied, where $\chi^2_{4\pi(\gamma)}$ is
$\chi^2$ of the kinematic fit.
It decreases the number of background $e^+e^- \to \pi^+\pi^-\pi^0\pi^0$ events by a factor of
10, rejecting about 40\% of the signal events.

\begin{figure}
  \includegraphics[width=0.47\textwidth]{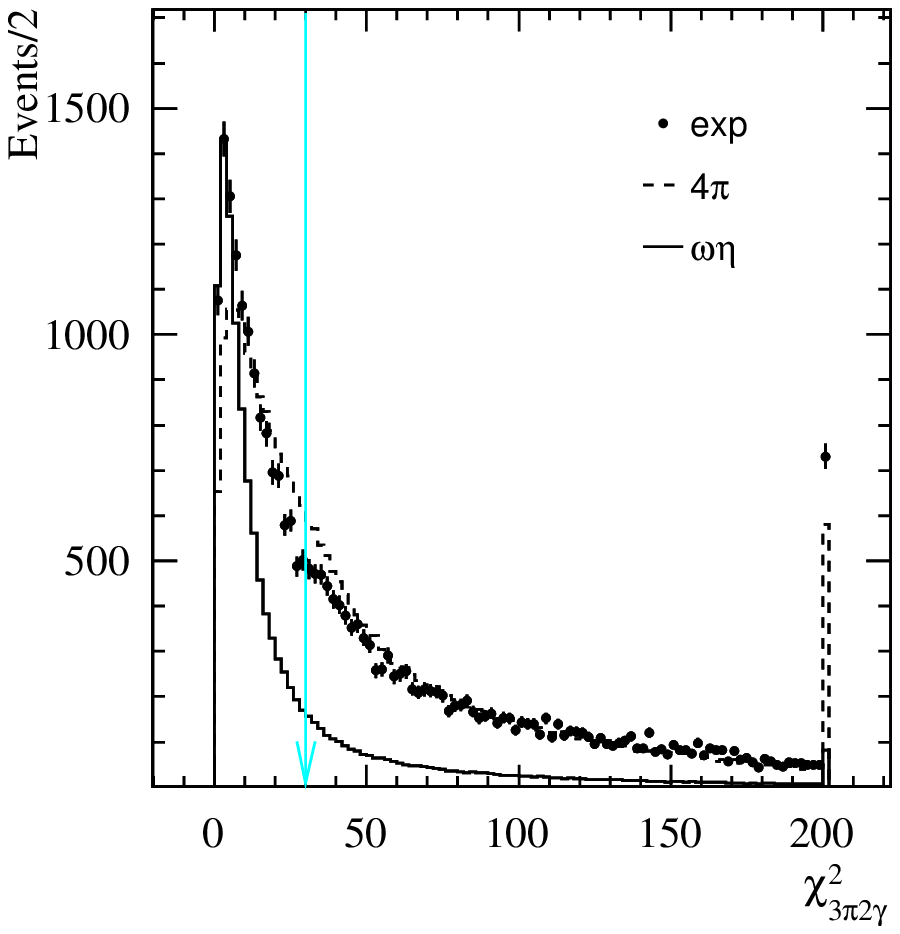}
  \hfill
  \includegraphics[width=0.47\textwidth]{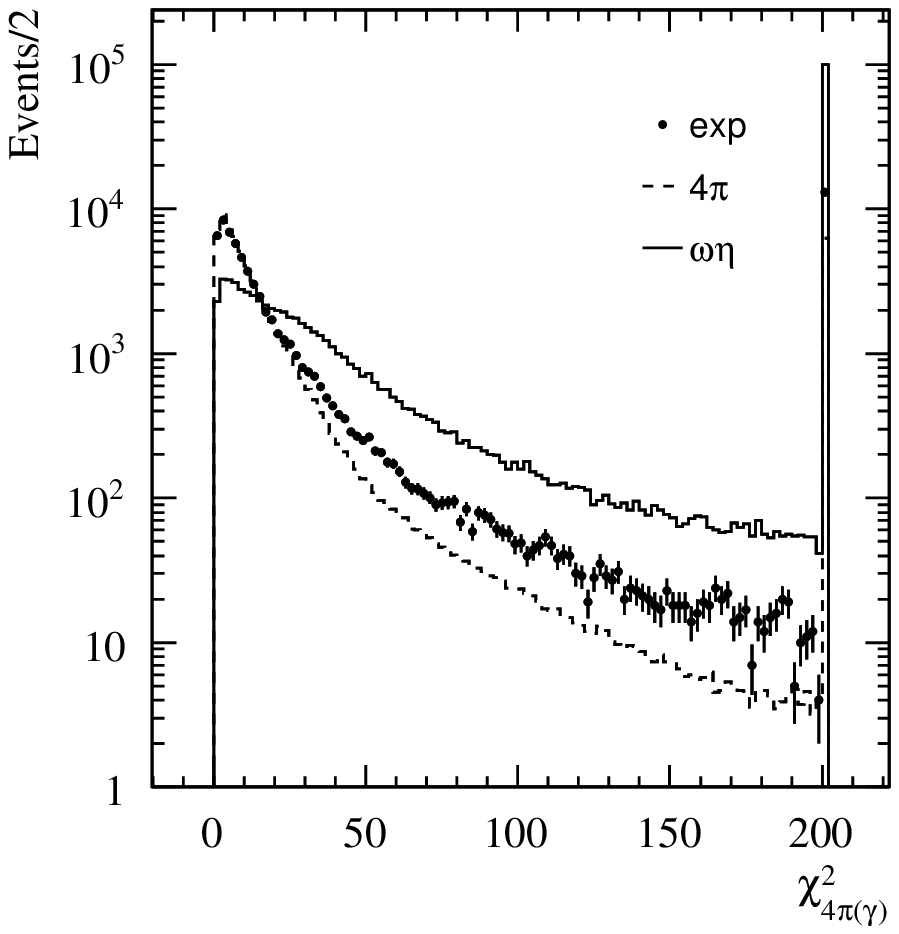}
  \caption{The $\chi^2_{3\pi2\gamma}$ distribution (left) and the $\chi^2_{4\pi(\gamma)}$
    distribution (right) for data events (points with error bars) and simulated $e^+e^- \to
    \omega\eta \to \pi^+\pi^-\pi^0\eta$ and $e^+e^- \to \pi^+\pi^-\pi^0\pi^0$ events (solid and
    dashed histograms).
    The arrow indicates the boundary of the condition $\chi^2_{3\pi2\gamma} < 30$.}
  \label{FIG:CHI2}
\end{figure}

The $\chi^2_{3\pi2\gamma}$ and $\chi^2_{4\pi(\gamma)}$ distributions for data events are shown
in Fig.~\ref{FIG:CHI2} in comparison with the distributions for simulated $e^+e^- \to
\omega\eta \to \pi^+\pi^-\pi^0\eta$ and $e^+e^- \to \pi^+\pi^-\pi^0\pi^0$ events.
The last bin of these distributions contains events with the $\chi^2$ value greater than 200.

%==============================================
\section{Determination of the number of signal events}
\label{FIT}

\begin{figure}
  \includegraphics[width=0.47\textwidth]{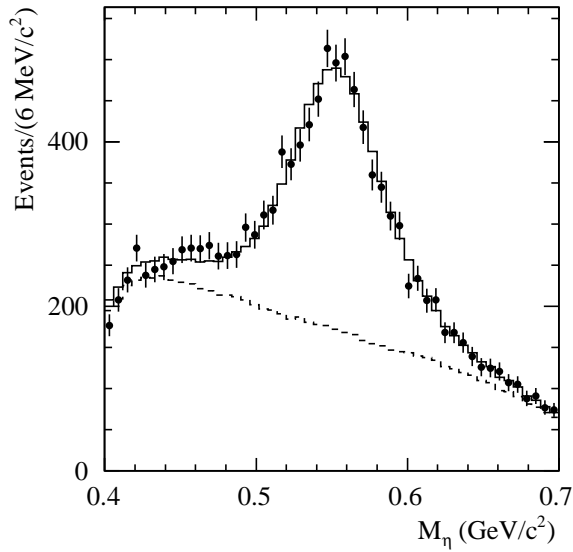}
  \caption{The $M_\eta$ spectrum for selected data events (points with error bars).
    The solid histogram is the result of the fit to the data spectrum with a sum of signal and
    background distributions.
    The fitted background contribution is shown by the dashed histogram.}
  \label{FIG:META}
\end{figure}

The $M_\eta$ spectrum for the selected 13113 data events is shown in Fig.~\ref{FIG:META}.
It is seen that only about 35\% of events contain an $\eta$ meson.
To determine the number of signal events the $M_\eta$ spectrum in each energy interval is
fitted with a sum of signal and background distributions.
The background distribution is obtained using simulation of the main background process $e^+e^-
\to \pi^+\pi^-\pi^0\pi^0$.
A possible inaccuracy in prediction of number background events is taken into account by
introducing a scale factor $\alpha_{4\pi}$.
\renewcommand{\var}{1.594~GeV}
For energies below \var, the value of $\alpha_{4\pi}$ found in the fit is consistent with unity.
At higher energies, there is significant background contribution from other processes, e.g.,
$e^+e^- \to \pi^+\pi^-\pi^0\pi^0\pi^0$.
In this region $\alpha_{4\pi}$ is fixed to unity, and a linear function is added to describe
contribution of other background processes.
It is worth noting that in the energy region above \var\ the shape of the $M_\eta$ distribution for
$e^+e^- \to \pi^+\pi^-\pi^0\pi^0$ events is close to linear.

The signal distribution is described by a sum of three Gaussian functions with parameters
determined from the fit to the simulated $M_\eta$ distribution obtained as a sum of the
distributions of the four subprocesses~(\ref{EQ:OMETA})--(\ref{EQ:NRES}).
To take into account a possible inaccuracy of the signal simulation, two parameters are
introduced: mass shift $\Delta M_\eta$ and $\Delta \sigma_{M_\eta}$.
The latter is quadratically added to all Gaussian sigmas.
These parameters are determined from the fit to the $M_\eta$ spectrum for events with $E \geq
1.544$~GeV and found to be $\Delta M_\eta = -0.9 \pm 1.0$~MeV/c$^2$ and $\Delta\sigma_{M_\eta} =
12.0 \pm 3$~MeV/c$^2$.

The number of signal and background events obtained from the fit to the $M_\eta$ spectrum in
Fig.~\ref{FIG:META} is $4643 \pm 126$ and $8519 \pm 185$, respectively.
The events with $\eta$ meson belong to the process $e^+e^- \to \pi^+\pi^-\pi^0\eta$.
The same fit is performed in the 13 energy intervals.
The obtained numbers of $e^+e^- \to \pi^+\pi^-\pi^0\eta$ events are listed in
Table~\ref{TAB:EXP}.

To estimate the systematic uncertainty in the number of signal events due to the imperfect
description of the shape of the background distribution, we perform the fit with an additional
linear background below \var, and without the linear background but with free $\alpha_{4\pi}$
above.
The uncertainty associated with the difference between data and simulation in the $\eta$-meson
line shape is estimated by variation of the parameters $\Delta M_\eta$ and $\Delta
\sigma_{M_\eta}$ within their errors.
The obtained systematic uncertainties are listed in Table~\ref{TAB:EXP}.

\section{Separation of intermediate states}
\label{FITCHAN}

\begin{figure}
  \includegraphics[width=0.47\textwidth]{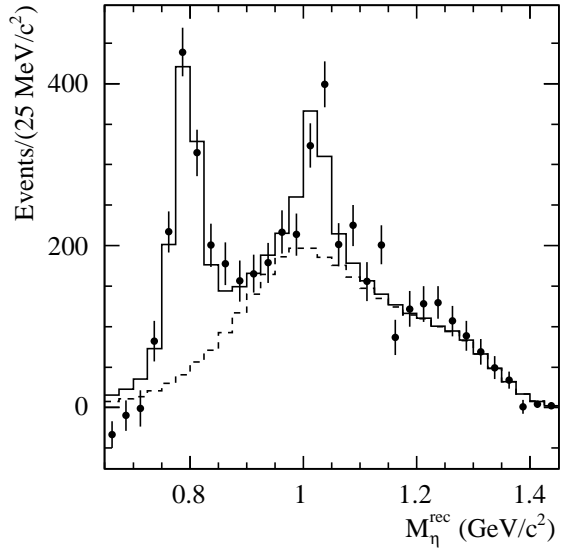}
  \caption{The $M_\eta^{rec}$ distribution for selected $e^+e^- \to \pi^+\pi^-\pi^0\eta$ data
    events (points with error bars).
    The solid histogram represents the result of the fit to the $M_\eta^{rec}$ distribution
    with the sum of simulated distributions for the processes~(\ref{EQ:OMETA})--(\ref{EQ:NRES}).
    The dashed histogram shows the fitted distribution for the sum of the
    processes~(\ref{EQ:ARHO}) and (\ref{EQ:NRES}).}
  \label{FIG:METAREC}
\end{figure}

The distribution of the invariant mass of the system recoiling against the $\eta$ meson
$M_\eta^{rec}$ for selected data events containing $\eta$ meson is shown in
Fig.~\ref{FIG:METAREC}.
The number of events in each $M_\eta^{rec}$ bin is determined from the fit to the $M_\eta$
distribution as described in Sec.~\ref{FIT}.
The peaks in the $M_\eta^{rec}$ spectrum at the $\omega$ and $\phi$ masses correspond to
$\omega\eta$ and $\phi\eta$ events.
The $a_0\rho$ and $nres$ intermediate states have wide $M_\eta^{rec}$ distributions with
similar shapes.

\renewcommand{\var}{1.844~GeV}
The contribution of the $a_0\rho$ mechanism can be observed in the $\eta\pi$ invariant mass
spectrum.
The $\eta\pi^0$ mass spectrum for data events containing $\eta$ meson from the energy region
$E \geq$~\var, where the $a_0\rho$ mechanism dominates~\cite{CMD}, is shown in
Fig.~\ref{FIG:ARHO}.
The spectrum is fitted by the sum of the $a_0\rho$, $nres$, and $\phi\eta$ distribution.
The number of $\phi\eta$ events and the absence of $\omega\eta$ events are determined from the
fit to the $M_\eta^{rec}$ distribution for the energy region $E \geq$~\var.
The clear signal of the $a_0^0$ meson is seen in Fig.~\ref{FIG:ARHO}.
It should be noted that the $a_0^0$ peak contains only one-third of $a_0\rho$ events.
Unfortunately, our $\eta\pi$-mass resolution is not sufficiently good to unambiguously separate
between $a_0\rho$ and $nres$ contributions below \var, where the fraction of $a_0\rho$
events is lower.
Therefore, in the further analysis we do not separate these two mechanisms.
Instead, we analyze $M_\eta^{rec}$ distributions at different c.m.\ energy intervals and
measure the $\omega\eta$ and $\phi\eta$ contributions, and the total contribution of the
$a_0\rho$ and $nres$ intermediate states.
\begin{figure}
  \includegraphics[width=0.47\textwidth]{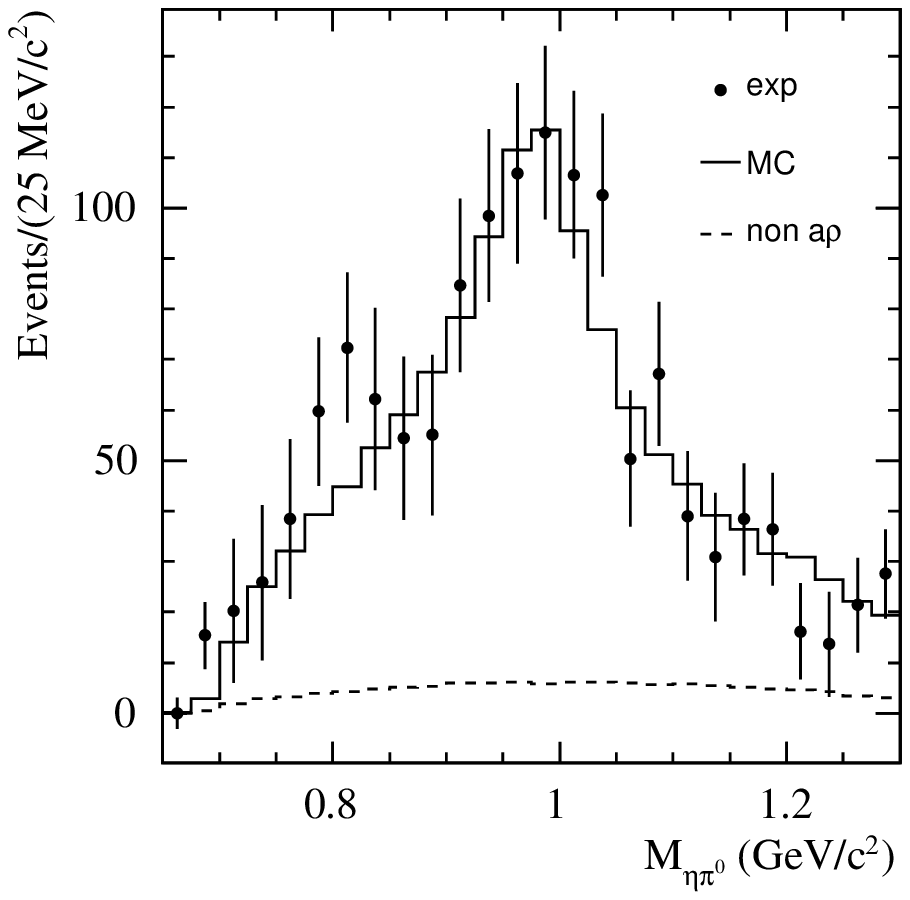}
  \caption{The distribution of $\eta\pi^0$ invariant mass for data events (points with error
    bars) with energy $E \geq$~\var.
    The solid histogram represents the result of the fit with the sum of simulated
    distributions for the $a_0\rho$, $\phi\eta$, and $nres$ intermediate states.
    The dashed histogram shows the fitted sum of the $\phi\eta$ and $nres$ distributions.}
  \label{FIG:ARHO}
\end{figure}

The $M_\eta^{rec}$ distributions are fitted with a sum of simulated distributions for the
intermediate states $\omega\eta$, $\phi\eta$, $a_0\rho$, and $nres$.
The $\omega\eta$ and $\phi\eta$ distributions are described by triple-Gaussian functions, the
parameters of which are determined from the fit to the simulated distributions.
The data-simulation difference in the $\omega$-meson line shape was studied in
Ref.~\cite{OMETA2} using $e^+e^- \to \omega\eta$ events.
The difference is parametrized by the two parameters $\Delta M_\omega$ and $\Delta
\sigma_{M_\omega}$.
The latter parameter was found consistent with zero, while $\Delta M_\omega = 7.5 \pm
1.9$~MeV/c$^2$.
The same correction is applied to the $\phi$-meson line shape.
The histograms obtained using MC simulation are used to describe the $a_0\rho$ and $nres$
distributions.
The uncertainties on $\Delta M_\omega$ and $\Delta M_\omega = \Delta M_\phi$ are used to
estimate the systematic uncertainty in the number of signal events for all contributions.
The free fit parameters are the numbers of $\omega\eta$ and $\phi\eta$ events, and the total
number of the $a_0\rho$ and $nres$ events.
The ratio of $a_0\rho$ to $nres$ events is fixed at the value measured in Ref.~\cite{CMD} and
is allowed to vary within its uncertainty during the fit.
The result of the fit to the $M_\eta^{rec}$ for all selected data events containing $\eta$
meson is represented in Fig.~\ref{FIG:METAREC}.

The $e^+e^- \to \omega\eta$ process was studied previously in the SND experiment in
Ref.~\cite{OMETA2}, using the same data sample.
The number of $\omega\eta$ events in the current analysis is about 15\% larger than that in
Ref.~\cite{OMETA2}.
One-third of this difference is due to the difference between the two analyses in the total
number of selected events containing $\eta$ meson.
In the current analysis we use more accurate $\eta$-meson line shape, which take into account
all four intermediate states contributing into the $e^+e^- \to \pi^+\pi^-\pi^0\eta$ reaction,
while in the previous the line shape was extracted from $e^+e^- \to \omega\eta$ simulation.
The remaining 10\% is associated with the shape of the non-$\omega\eta$ distribution, which was
described in Ref.~\cite{OMETA2} by a linear function.
The $a_0\rho+nres$ model used in this analysis is certainly more realistic.
In particular, it describes well the $M_\eta^{rec}$ spectrum in the full range of $M_\eta^{rec}$
variation (see Fig.~\ref{FIG:METAREC}).
On the other hand, we observe that the measurement of the $e^+e^- \to \omega\eta$ cross section
is very sensitive to the shape of the non-$\omega\eta$ $M_\eta^{rec}$ distribution.
Therefore, an additional systematic uncertainty on the number of fitted $\omega\eta$ events is
introduced to account this sensitivity, which is estimated to be 10\%, the difference between
results obtained with the linear background shape and the $a_0\rho+nres$ model.
The same uncertainty is assigned to the number of fitted $\phi\eta$ events.
The $\omega\eta$ and $\phi\eta$ systematic uncertainties are translated to the uncertainty on
the number of $a_0\rho+nres$ events as $\sqrt{\Delta N^2_{\omega\eta}+\Delta N^2_{\phi\eta}}$.

%==============================================
\section{Detection efficiency}

The MC simulation used in this analysis takes into account radiative corrections, in particular
photon emission from the initial state.
Therefore, the detection efficiency ($\varepsilon_{\rm MC}$) depends on the Born cross section
for the simulated process and is determined using the iterative procedure described in
Sec.~\ref{DATA}.
In our case of the resonance energy region, the detection efficiency $\varepsilon_{\rm MC}$ may
strongly deviate, due to the initial state radiation, from the efficiency $\varepsilon_0$
determined at $E_\gamma = 0$, where $E_\gamma$ is the energy of the photon emitted from the
initial state.

\begin{figure}
  \includegraphics[width=0.47\textwidth]{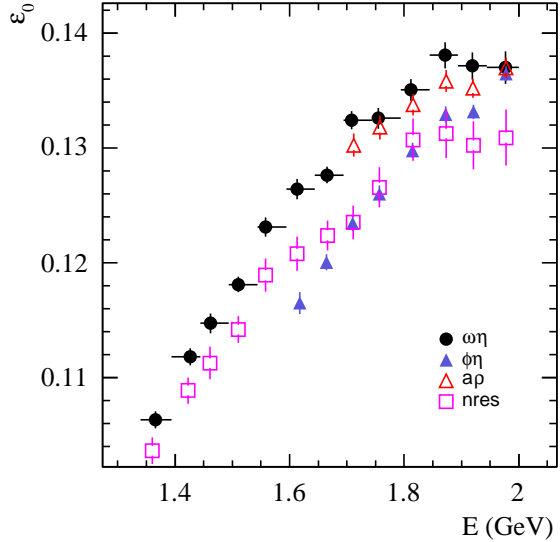}
  \caption{The $\varepsilon_0$ energy dependences for the four intermediate states contributing
    to the process $e^+e^- \to \pi^+\pi^-\pi^0\eta$.}
  \label{FIG:EFF0}
\end{figure}

The dependence of the efficiency on $E_\gamma$ may be parametrized as
$\varepsilon_0(E)r(E,E_\gamma)$.
The shape of the function $r(E,E_\gamma)$ is defined by the cut on $\chi^2_{3\pi2\gamma}$ and
is practically the same for all processes under study.
The energy dependences of $\varepsilon_0$ for the four intermediate states contributing to the
process $e^+e^- \to \pi^+\pi^-\pi^0\eta$ are shown in Fig.~\ref{FIG:EFF0}.
It is seen that in the energy range under study the efficiencies change from 10 to 14\%.
However, for the given energy the differences between the four efficiencies are within 5\%.
This difference is taken as an estimate of the additional model uncertainty on the detection
efficiency in the measurements of the total $e^+e^- \to \pi^+\pi^-\pi^0\eta$ cross section and
the cross section for the sum of the $a_0\rho$ and $nres$ intermediate states.
For these measurements the simulation in the $e^+e^- \to \rho(1450)\pi$, $\rho(1450) \to
\rho\eta$ model is used.
The efficiency obtained in this model is corrected with a scale factor of
$\bar{\varepsilon}_0(E)/\varepsilon_0^{nres}(E)$, where $\varepsilon_0^{nres}$ is the
efficiency at $E_\gamma=0$ obtained in the $e^+e^- \to \rho(1450)\pi$ model, and
$\bar{\varepsilon}_0$ is the efficiency averaged over the four (two) intermediate states for
the total ($a_0\rho+nres$) cross section measurement with the weights equal to their expected
fractions in the number of selected events.
\begin{figure}
  \includegraphics[width=0.47\textwidth]{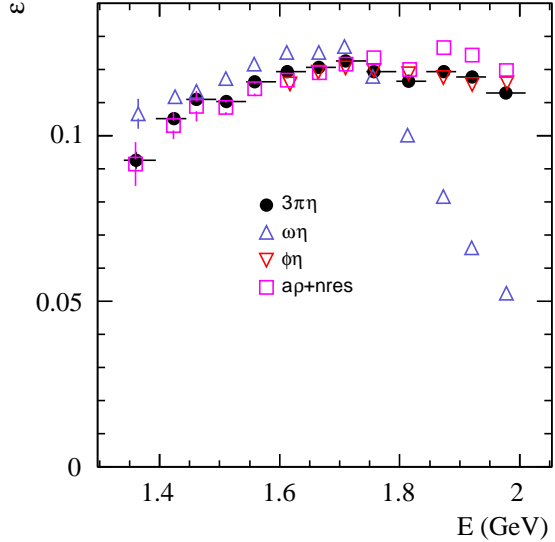}
  \caption{The energy dependence of the detection efficiency, corrected as described in text,
    for the process $e^+e^- \to 3\pi\eta$ and the intermediate states $\omega\eta$,
    $\phi\eta$, and $a_0\rho+nres$.}
  \label{FIG:EFF}
\end{figure}

\begin{table*}
  \caption{The weighted average energy ($\protect\widebar{E}$\,) for the c.m.\ energy interval
    specified, integrated luminosity ($L$), number of selected $\pi^+\pi^-\pi^0\eta$ data
    events ($N$), the radiative correction factor ($1+\delta$) and detection efficiency
    ($\varepsilon$) for the process $e^+e^- \to \pi^+\pi^-\pi^0\eta$.
    For $N$ and $\varepsilon$ the statistical and systematic errors are quoted.
    For $1+\delta$ the systematic errors are quoted.}
  \label{TAB:EXP}
  \begin{ruledtabular}
  \begin{tabular}{*{5}c}
    $\widebar{E}$ (GeV) & $L$ (nb$^{-1}$) & $N$		& $1+\delta$ & $\varepsilon$ (\%)\\
    \hline
    1.361 [1.340--1.394) & 2082 & $-37 \pm 15 \pm 7 $ & $0.81 \pm 0.00$ & $9.3  \pm 0.3 \pm 0.6$\\
    1.424 [1.394--1.444) & 2256 & $35  \pm 19 \pm 16$ & $0.86 \pm 0.01$ & $10.5 \pm 0.2 \pm 0.7$\\
    1.461 [1.444--1.494) & 1095 & $38  \pm 16 \pm 4 $ & $0.86 \pm 0.01$ & $11.1 \pm 0.2 \pm 0.8$\\
    1.511 [1.494--1.544) & 2193 & $146 \pm 27 \pm 16$ & $0.86 \pm 0.01$ & $11.0 \pm 0.1 \pm 0.8$\\
    1.558 [1.544--1.594) & 1024 & $95  \pm 19 \pm 2 $ & $0.85 \pm 0.01$ & $11.6 \pm 0.1 \pm 0.8$\\
    1.613 [1.594--1.644) & 1008 & $246 \pm 26 \pm 7 $ & $0.84 \pm 0.00$ & $11.9 \pm 0.1 \pm 0.8$\\
    1.665 [1.644--1.694) & 1854 & $941 \pm 44 \pm 11$ & $0.86 \pm 0.01$ & $12.1 \pm 0.0 \pm 0.8$\\
    1.710 [1.694--1.744) & 1540 & $703 \pm 40 \pm 11$ & $0.93 \pm 0.00$ & $12.3 \pm 0.0 \pm 0.8$\\
    1.756 [1.744--1.794) & 1722 & $538 \pm 40 \pm 11$ & $1.00 \pm 0.01$ & $11.9 \pm 0.1 \pm 0.8$\\
    1.815 [1.794--1.844) & 2929 & $614 \pm 47 \pm 21$ & $1.05 \pm 0.01$ & $11.7 \pm 0.0 \pm 0.8$\\
    1.873 [1.844--1.894) & 2678 & $511 \pm 43 \pm 22$ & $1.05 \pm 0.03$ & $11.9 \pm 0.1 \pm 0.8$\\
    1.921 [1.894--1.944) & 3702 & $596 \pm 48 \pm 23$ & $1.06 \pm 0.03$ & $11.8 \pm 0.0 \pm 0.8$\\
    1.977 [1.944--2.000) & 2930 & $343 \pm 42 \pm 21$ & $1.15 \pm 0.10$ & $11.3 \pm 0.1 \pm 0.8$
  \end{tabular}
  \end{ruledtabular}
\end{table*}

The detection efficiencies obtained using MC simulation are shown in Fig.~\ref{FIG:EFF}.
The steep decrease of the $e^+e^- \to \omega\eta$ efficiency is explained by a very small value
of the Born cross section above 1.8~GeV.
In this energy region practically all detected $\omega\eta$ events contain a photon emitted
from the initial state, which distorts the event kinematics.

Imperfect simulation of detector response leads to a difference between the actual detection
efficiency $\varepsilon$ and the efficiency determined using MC simulation $\varepsilon_{\rm
MC}$
\begin{equation}
  \varepsilon = \varepsilon_{\rm MC}\prod_{i=1}^{n}(1+\kappa_i),
  \label{EQ:EFF}
\end{equation}
where $\kappa_i$ are the efficiency corrections for different effects.
The main selection criterion for signal events $\chi^2_{3\pi2\gamma} < 30$ is mostly affected
by uncertainties in simulation.
The quality of the simulation of the $\chi^2_{3\pi2\gamma}$ distribution is studied in
Ref.~\cite{OMETA2} using $e^+e^- \to \omega\pi^0 \to \pi^+\pi^-\pi^0\pi^0$ events, which has a
large cross section in the energy region under study and the same number of the final particles
as the process under study.
The correction is found to be $\kappa_1 = (2.5 \pm 1.1)$\%.
The correction on the $\chi^2_{4\pi(\gamma)} > 200$ condition was also studied in
Ref.~\cite{OMETA2} and was found to be consistent with zero with a systematic uncertainty of 4.6\%.

The difference between data and simulation in photon conversion in detector material before the
tracking system is studied using events of the process $e^+e^- \to \gamma\gamma$.
The corresponding efficiency correction is $\kappa_2 = (-1.35 \pm 0.05)$\%.
The largest part of the systematic uncertainties associated with data-MC simulation difference
in track loss cancels as a result of luminosity normalization.
The difference in the track reconstruction for electrons and pions was studied in
Ref.~\cite{TRACKS}.
The corresponding correction $\kappa_3 = (-0.3 \pm 0.2)$\%.

The total correction is $\kappa = (0.9 \pm 4.7)$\%.
The corrected values of the detection efficiency for the process $e^+e^- \to
\pi^+\pi^-\pi^0\eta$ are listed in Table~\ref{TAB:EXP}.
For the processes $e^+e^- \to \omega\eta$ and $\phi\eta$, the systematic error is 4.8\% and
includes the uncertainty of the efficiency correction (4.7\%), the uncertainty of the iterative
procedure described in Sec.~\ref{DATA}.
For $e^+e^- \to \pi^+\pi^-\pi^0\eta$ and $e^+e^- \to a_0\rho+nres$, the 5\% uncertainty related
to inaccurately known process dynamics is added, and the total uncertainty is 6.9\%.

%==============================================
\section{Determination of the Born cross sections}
\label{CRS}

The experimental values of the visible cross section for the processes under study are
calculated as follows:
\begin{equation}
  \sigma_{vis,i} = \frac{N_i}{L_i \varepsilon_i B},
  \label{EQ:CRSEXP}
\end{equation}
where $N_i$, $L_i$, and $\varepsilon_i$ are the number of selected data events, integrated
luminosity, and detection efficiency for the $i$th energy interval.
The parameter $B$ is equal to the branching fraction $B(\omega \to \pi^+\pi^-\pi^0) = 0.892 \pm
0.007$~\cite{PDG} for $e^+e^- \to \omega\eta$, $B(\phi \to \pi^+\pi^-\pi^0) = 0.1524 \pm 0.0033$
for $e^+e^- \to \phi\eta$, and unity for others.

The visible cross section ($\sigma_{vis}$) is related to the Born cross section ($\sigma$) by
the following expression~\cite{KURAEV}:
\begin{equation}
  \sigma_{vis}(E) = \int_0^{x_{max}} F(x,E)\sigma(E\sqrt{1-x})dx,
  \label{EQ:CRSVIS}
\end{equation}
where $x$ is the beam-energy fraction carried away by photons emitted from the initial state,
the function $F(x,E)$ describes the probability of radiation of photons with total energy
$xE/2$, and $x_{max}=1-(2m_{\pi^+}+m_{\pi^0}+m_\eta)^2/E^2$.
The right side of Eq.~(\ref{EQ:CRSVIS}) can be rewritten in the more conventional form
\begin{equation}
  \int_0^{x_{max}} F(x,E)\sigma(E\sqrt{1-x})dx = \sigma(E)(1+\delta(E)),
  \label{EQ:CRSVIS2}
\end{equation}
where $\delta(E)$ is the radiative correction.

Experimental values of the Born cross section are obtained as follows.
The energy dependence of the measured visible cross section is fitted with
Eq.~(\ref{EQ:CRSVIS}), in which the Born cross section is given by a theoretical model
describing data well.
The model parameters obtained in the fit are used to calculate $\delta(\widebar{E}_i)$ using
Eq.~(\ref{EQ:CRSVIS2}).
Here $\widebar{E}_i$ is the weighted average c.m.\ energy for $i$th energy interval.
The values of the Born cross section are then obtained using the equation
\begin{equation}
  \sigma_i = \sigma_{vis,i}/(1+\delta(\widebar{E}_i)).
  \label{EQ:CRS}
\end{equation}

\begin{figure*}
  \includegraphics[width=0.47\textwidth]{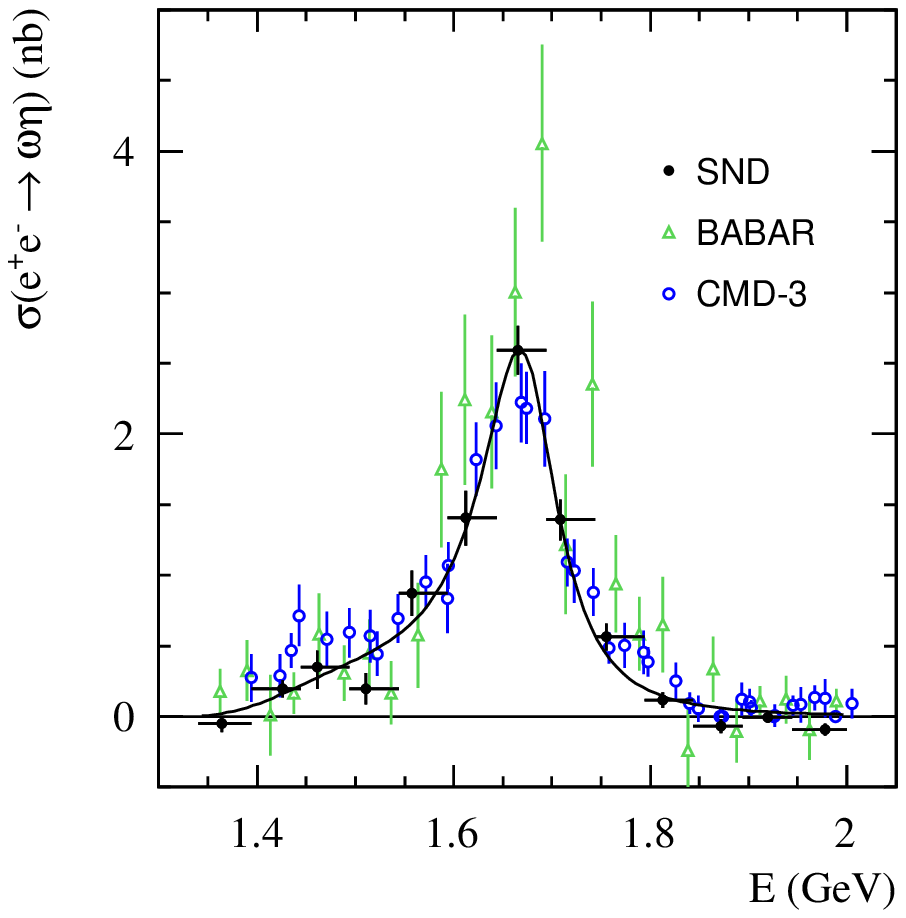}
  \put(-150,150){\textbf{\large (a)}} \hfill
  \includegraphics[width=0.47\textwidth]{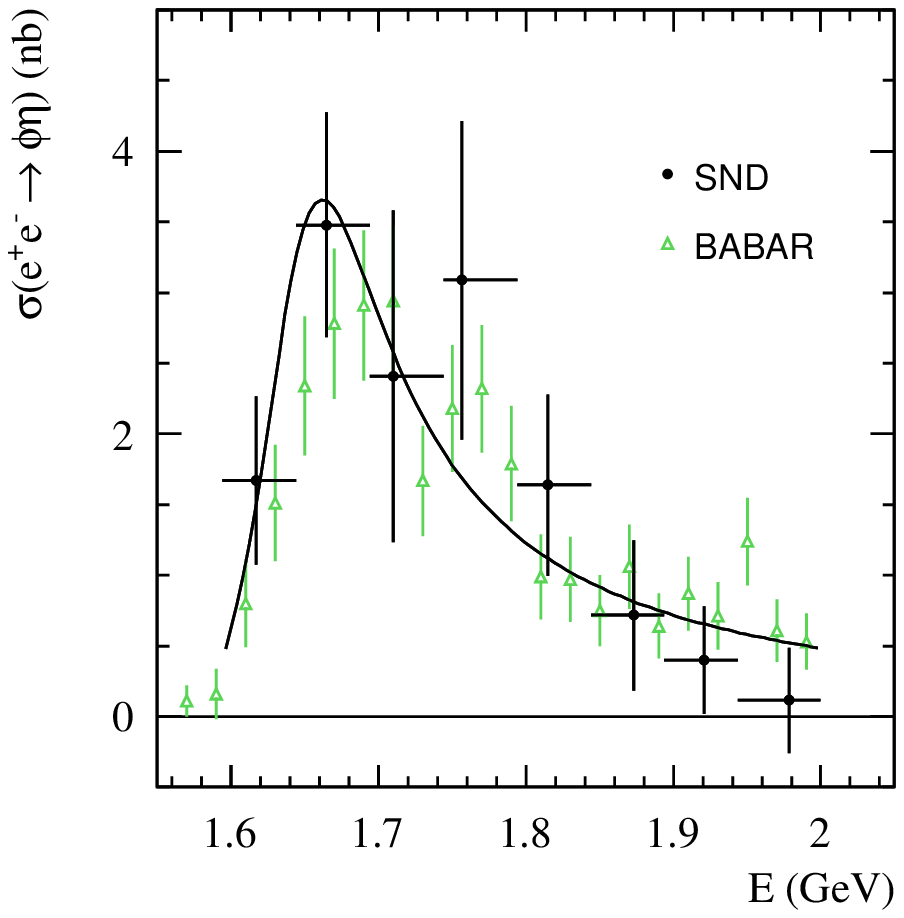}
  \put(-60,120){\textbf{\large (b)}}\\
  \includegraphics[width=0.47\textwidth]{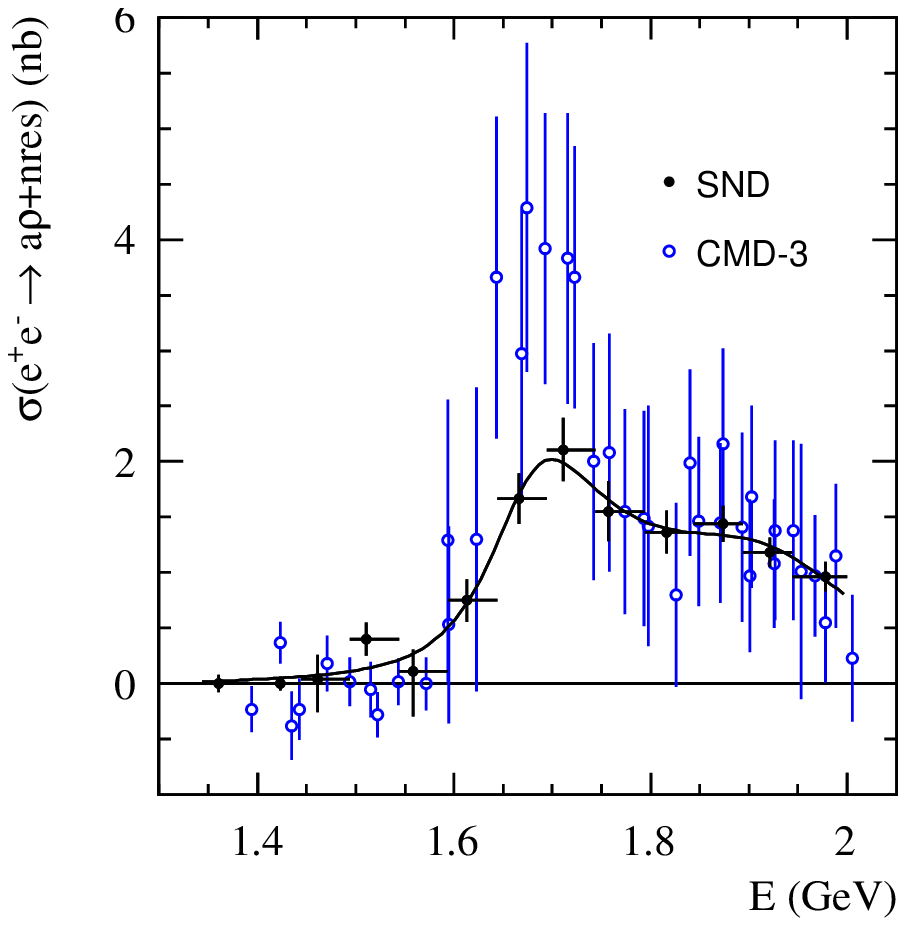}
  \put(-150,150){\textbf{\large (c)}} \hfill
  \includegraphics[width=0.47\textwidth]{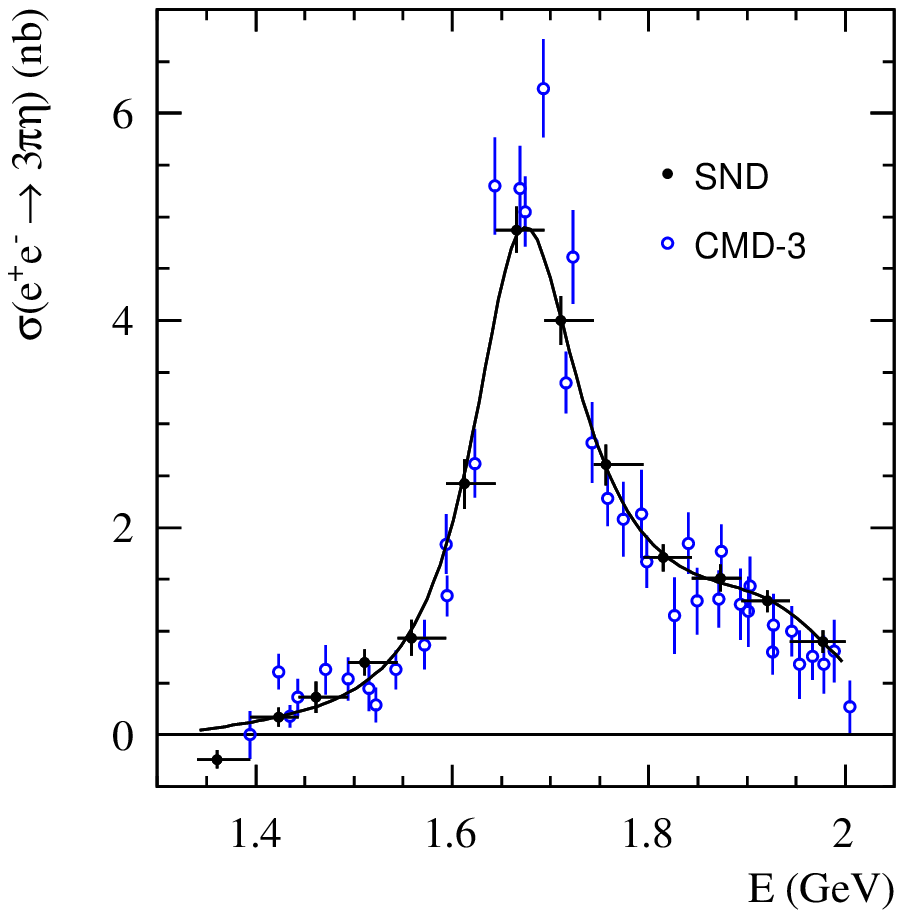}
  \put(-150,150){\textbf{\large (d)}}
  \caption{The Born cross sections for processes $e^+e^- \to \omega \eta$ (a),
    $e^+e^- \to \phi \eta$ (b), $e^+e^- \to a_0\rho+nres \to \pi^+\pi^-\pi^0\eta$ (c), and
    $e^+e^- \to \pi^+\pi^-\pi^0\eta$ (d) measured in this work (filled circles) and in the
    previous experiments~\cite{OMETA,PHIETA2,CMD} (open circles and triangles).
    Only statistical errors are drawn.
    The curves are the results of the fit described in the text.}
  \label{FIG:CRS}
\end{figure*}

The two-resonance model is used to parametrize the Born cross sections
\begin{widetext}
  \begin{equation}
    \sigma(E) = \frac{12\pi}{E^3}\left|
    \sqrt{\frac{B_{V'}}
      {P_f(m_{V'})}}\frac{m_{V'}^{3/2}\Gamma_{V'}}{D_{V'}}+
    \sqrt{\frac{B_{V''}}
      {P_f(m_{V''})}}\frac{m_{V''}^{3/2}\Gamma_{V''}}{D_{V''}}e^{i\varphi}
    \right|^2P_f(E),
    \label{EQ:VMD}
  \end{equation}
\end{widetext}
where $m_V$ and $\Gamma_V$ are the mass and width of the resonance $V$ ($V=V'$ or $V''$),
$D_V = E^2-m^2_V+iE\Gamma_V$,
$B_V = B(V \to e^+e^-)B(V \to f)$ is the product of the branching fractions for the $V$ decay
to $e^+e^-$ and the final state $f$, and $P_f(E)$ is the phase space factor.
The obtained values of the Born cross sections are shown in Fig.~\ref{FIG:CRS} together with
the fitted curves.

For $e^+e^- \to \omega\eta$, the first term in Eq.~(\ref{EQ:VMD}) is associated with the
$\omega(1420)$ resonance, the second is a sum of contributions of the $\omega(1650)$ and
$\phi(1680)$ resonances, and $P_f(E)=q^3_\omega(E)$, where $q_\omega(E)$ is the $\omega$
momentum in the reaction $e^+e^- \to \omega\eta$.
The phase between the first and second terms in Eq.~(\ref{EQ:VMD}) is chosen to be equal
to $\pi$~\cite{OMETA2}.
In the fit to the $e^+e^- \to \omega\eta$ cross-section data, the free parameters are $B_{V'}$,
$B_{V''}$, $m_{V''}$, and $\Gamma_{V''}$.
The $V'$ mass and width are fixed at the Particle Data Group (PDG) values for
$\omega(1420)$~\cite{PDG}.
The obtained fit parameters are listed in Table~\ref{TAB:APPROX}.
The fitted $V''$ mass is in agreement with the PDG mass of both $\omega(1650)$ and $\phi(1680)$
resonances~\cite{PDG}, while the fitted width is smaller than the PDG estimate for the
$\omega(1650)$ width, $315 \pm 35$~MeV/c$^2$~\cite{PDG}, but agrees with the PDG value, $150 \pm
50$~MeV/c$^2$~\cite{PDG}, for the $\phi(1680)$ resonance.
\begin{table}
  \caption{The fit parameters for the processes $e^+e^- \to \omega\eta$ and $e^+e^- \to
    \phi\eta$, and $\chi^2/\nu$, where $\nu$ is the number degrees of freedom.}
  \label{TAB:APPROX}
  \begin{ruledtabular}
  \begin{tabular}{lcc}
      				& $\omega\eta$		 & $\phi\eta$\\
    \hline
    $B_{V'} \times 10^{7}$	& $0.21^{+0.10}_{-0.08}$ & ---\\
    $B_{V''} \times 10^{7}$	& $5.62^{+0.45}_{-0.42}$ & $5.64^{+1.74}_{-1.80}$\\
    $M_{V''}$ (MeV)		& $1673^{+6}_{-7}$       & $1641^{+24}_{-18}$\\
    $\Gamma_{V''}$ (MeV)	& $95 \pm 11$		 & $103^{+26}_{-24}$\\
    $\chi^2/\nu$		& $10.5/9$	         & $4.4/5$
  \end{tabular}
  \end{ruledtabular}
\end{table}

The $\phi\eta$ data are well described by a single-resonance model with $B_{V'}=0$ and
$P_f(E)=q^3_\phi(E)$, where $q_\phi(E)$ is the $\phi$ momentum in the reaction $e^+e^- \to
\phi\eta$.
The obtained $V''$ mass and width listed in Table~\ref{TAB:APPROX} are in agreement with the
PDG values for the $\phi(1680)$.

The $e^+e^- \to a_0\rho+nres \to \pi^+\pi^-\pi^0\eta$ and $e^+e^- \to \pi^+\pi^-\pi^0\eta$ cross
sections are described by the model (\ref{EQ:VMD}) with seven free parameters ($B_{V'}$,
$m_{V'}$, $\Gamma_{V'}$, $B_{V''}$, $m_{V''}$, $\Gamma_{V''}$, and $\varphi$) and
$P_f(E)=q_\omega(E)$.
This model has no physical sense, but describes data well, and therefore can be used to
calculate radiative corrections.

The obtained values of the Born cross sections are listed in Table~\ref{TAB:CRS}.
The numerical values of radiative correction for the total $e^+e^- \to \pi^+\pi^-\pi^0\eta$ are
listed in Table~\ref{TAB:EXP}.
The uncertainty on the radiative correction is estimated by varying the fitted parameters
within their errors.
For the cross sections we quote the statistical and systematic errors.
The latter includes the uncertainties on the detection efficiency (systematic and model),
number of selected events, luminosity, and radiative correction.
The energy-independent (correlated) part of the uncertainty for the total $e^+e^- \to
\pi^+\pi^-\pi^0\eta$ cross section is 7\%.

\begin{table*}
  \caption{The measured Born cross sections for the processes $e^+e^- \to \pi^+\pi^-\pi^0\eta$
    ($\sigma$), $e^+e^- \to \omega\eta$ ($\sigma(\omega\eta)$), $e^+e^- \to \phi\eta$
    ($\sigma(\phi\eta)$), and the sum of processes $e^+e^- \to a_0\rho$ and $e^+e^- \to nres$
    ($\sigma(a_0\rho+nres)$).
    The statistical and systematic errors are quoted.}
  \label{TAB:CRS}
  \begin{ruledtabular}
  \begin{tabular}{*{5}c}
    $\widebar{E}$ (GeV) & $\sigma$ (nb) & $\sigma(\omega\eta)$ (nb) & $\sigma(\phi\eta)$ (nb) & $\sigma(a_0\rho+nres)$ (nb)\\
    \hline
    1.361 & $-0.24 \pm 0.09 \pm 0.05$ & $-0.05 \pm 0.06 \pm 0.02$ & ---			     & $0.00 \pm 0.08 \pm 0.00$\\
    1.424 & $0.17  \pm 0.09 \pm 0.08$ & $0.20  \pm 0.06 \pm 0.08$ & ---			     & $0.00 \pm 0.06 \pm 0.02$\\
    1.461 & $0.36  \pm 0.15 \pm 0.04$ & $0.35  \pm 0.15 \pm 0.05$ & ---			     & $0.04 \pm 0.29 \pm 0.03$\\
    1.511 & $0.70  \pm 0.13 \pm 0.09$ & $0.20  \pm 0.11 \pm 0.02$ & ---			     & $0.40 \pm 0.15 \pm 0.04$\\
    1.558 & $0.94  \pm 0.17 \pm 0.07$ & $0.87  \pm 0.16 \pm 0.12$ & ---			     & $0.11 \pm 0.41 \pm 0.09$\\
    1.613 & $2.42  \pm 0.24 \pm 0.18$ & $1.41  \pm 0.19 \pm 0.17$ & $1.67 \pm 0.60 \pm 0.20$ & $0.75 \pm 0.19 \pm 0.15$\\
    1.665 & $4.87  \pm 0.22 \pm 0.35$ & $2.59  \pm 0.17 \pm 0.30$ & $3.48 \pm 0.79 \pm 0.54$ & $1.67 \pm 0.22 \pm 0.29$\\
    1.710 & $4.00  \pm 0.24 \pm 0.28$ & $1.39  \pm 0.14 \pm 0.18$ & $2.41 \pm 1.17 \pm 0.32$ & $2.11 \pm 0.28 \pm 0.24$\\
    1.756 & $2.61  \pm 0.19 \pm 0.19$ & $0.56  \pm 0.10 \pm 0.07$ & $3.09 \pm 1.13 \pm 0.40$ & $1.55 \pm 0.27 \pm 0.14$\\
    1.815 & $1.71  \pm 0.13 \pm 0.14$ & $0.12  \pm 0.06 \pm 0.02$ & $1.64 \pm 0.64 \pm 0.25$ & $1.36 \pm 0.19 \pm 0.11$\\
    1.873 & $1.52  \pm 0.13 \pm 0.13$ & $-0.07 \pm 0.05 \pm 0.01$ & $0.72 \pm 0.53 \pm 0.09$ & $1.44 \pm 0.17 \pm 0.11$\\
    1.921 & $1.29  \pm 0.10 \pm 0.11$ & $-0.00 \pm 0.04 \pm 0.00$ & $0.40 \pm 0.38 \pm 0.06$ & $1.18 \pm 0.13 \pm 0.09$\\
    1.977 & $0.90  \pm 0.11 \pm 0.11$ & $-0.09 \pm 0.04 \pm 0.03$ & $0.12 \pm 0.37 \pm 0.02$ & $0.96 \pm 0.13 \pm 0.08$\\
  \end{tabular}
  \end{ruledtabular}
\end{table*}

In Fig.~\ref{FIG:CRS} our results on the cross sections are compared with the previous
measurements.
The obtained $e^+e^- \to \omega\eta$ cross section agrees with the CMD-3 measurement~\cite{CMD}.
Both the SND and CMD-3 results lie below the BABAR data~\cite{OMETA}.
The SND and BABAR~\cite{PHIETA} measurements of the $e^+e^- \to \phi\eta$ cross sections are in
reasonable agreement.
The significant difference between the SND and CMD-3 measurements is observed for the
$e^+e^- \to a_0\rho + nres$ cross sections.
The total $e^+e^- \to \pi^+\pi^-\pi^0\eta$ cross section measured by SND is, in general,
consistent with the CMD-3 result~\cite{CMD}.
The $\sim 15$\% difference in the cross section maximum is within the systematic uncertainties,
which are 7\% for SND and 11\% for CMD-3.

%==============================================
\section{Summary}

In the experiment with the SND detector at the VEPP-2000 $e^+e^-$ collider in the energy range
1.34--2.00~GeV the cross sections for the process $e^+e^- \to \pi^+\pi^-\pi^0\eta$ and its
subprocesses $e^+e^- \to \omega\eta$, $e^+e^- \to \phi\eta$ and $e^+e^- \to a_0\rho+nres$,
where $nres$ is the structureless $\pi^+\pi^-\pi^0\eta$ state, are measured.
The cross sections have a peak near the energy $\simeq 1650$~MeV.
The maximum value of the $e^+e^- \to \pi^+\pi^-\pi^0\eta$ cross section is 5~nb, about 8\% of
the total hadron cross section at this energy.
The result on the $e^+e^- \to \omega\eta$ cross section supersedes the previous SND
measurement~\cite{OMETA2}.
The obtained $e^+e^- \to \omega\eta$ and $e^+e^- \to \phi\eta$ cross sections are well fitted
in the VMD model with the $\omega(1420)$, $\omega(1650)$ and $\phi(1680)$ resonances.

%==============================================


\begin{thebibliography}{90}
  \bibitem{VEPP} A.~Romanov \emph{et al.},
    in Proceedings of PAC, 2013, Pasadena, CA, USA, p.\ 14.
  \bibitem{SND} M.~N.~Achasov \emph{et al.},
    Nucl.\ Instrum.\ Methods Phys.\ Res., Sect.\ A 598, 31 (2009);
    V.~M.~Aulchenko \emph{et al.}, ibid.\ 598, 102 (2009);
    A.~Yu.~Barnyakov \emph{et al.}, ibid.\ 598, 163 (2009);
    V.~M.~Aulchenko \emph{et al.}, ibid.\ 598, 340 (2009).
  \bibitem{MESON2016} V.~P.~Druzhinin \emph{et al.} (SND collaboration),
    %Measurement of hadron cross sections with the SND detector,
    EPJ Web Conf.\ \textbf{130}, 05004 (2016) [arXiv:1609.01040 [hep-ex]].
  \bibitem{CMD} R.~R.~Akhmetshin \emph{et al.} (CMD-3 Collaboration),
    %Study of the process $e^+e^- \to \pi^+\pi^-\pi^0\eta$ in the c.m.\ energy
    %range 1394--2005~MeV with the CMD-3 detector,
    Phys.\ Lett.\ B \textbf{773}, 150(2017) [arXiv:1706.06267 [hep-ex]].
  \bibitem{OMETA} B.~Aubert \emph{et al.} (BABAR Collaboration),
    %The $e^+e^- \to 3(\pi^+\pi^-), 2(\pi^+\pi^-\pi^0)$ and $K^+K^-2(\pi^+\pi^-)$
    %cross sections at center-of-mass energies from production threshold to 4.5~GeV measured
    %with initial-state radiation,
    Phys.\ Rev.\ D \textbf{73}, 052003 (2006).
  \bibitem{OMETA2} M.~N.~Achasov \emph{et al.} (SND collaboration),
    %Measurement of the $e^+e^- \to \omega\eta$ cross section below $\sqrt{s}=2$~GeV with the SND
    %detector,
    Phys.\ Rev.\ D \textbf{94}, 092002 (2016).
  \bibitem{PHIETA} B.~Aubert \emph{et al.} (BABAR Collaboration),
    %The $e^+e^- \to 2(\pi^+\pi^-)\pi^0, 2(\pi^+\pi^-)\eta, K^+K^-\pi^+\pi^-\pi^0 and
    %K^+K^-\pi^+\pi^-\eta Cross Sections Measured with Initial-State Radiation
    Phys.\ Rev.\ D \textbf{76}, 092005 (2007).
  \bibitem{PHIETA2} B.~Aubert \emph{et al.} (BABAR Collaboration),
    %Measurements of $e^{+} e^{-} \to K^{+} K^{-} \eta$, $K^{+} K^{-} \pi^0$ and $K^0_{s}
    %K^\pm \pi^\mp$ cross- sections using initial state radiation events,
    Phys.\ Rev.\ D \textbf{77}, 092002 (2008).
  \bibitem{PHIETA3} J.~P.~Lees \emph{et al.} (BABAR Collaboration),
    %Cross sections for the reactions $e^+ e^- \to K^0_S K^0_L\pi^0$, $K^0_S K^0_L\eta$, and
    %$K^0_S K^0_L\pi^0\pi^0$ from events with initial-state radiation,
    Phys.\ Rev.\ D \textbf{95}, 052001 (2017).
  \bibitem{PHIETA4} M.~N.~Achasov \emph{et al.} (SND collaboration),
    %Measurement of the $e^+e^- \to \eta K^+K^-$ cross section by means of the
    %SND Detector,
    Phys.\ Atom.\ Nucl.\ \textbf{81}, 205 (2018)
    [Yad.\ Fiz.\ \textbf{81}, no. 2, 195 (2018)].
  \bibitem{BEAM} D.~N.~Shemyakin \emph{et al.} (CMD-3 Collaboration),
    %Measurement of the $e^+e^- \to K^+K^-\pi^+\pi^-$ cross section with the CMD-3
    %detector at the VEPP-2000 collider,
    Phys.\ Lett.\ B \textbf{756}, 153 (2016).
  \bibitem{4PI} H.~Czyz, J.~H.~Kuhn, and A.~Wapienik,
    %Four-pion production in $\tau$ decays and $e^+e^-$ annihilation: An update
    Phys.\ Rev.\ D \textbf{77}, 114005 (2008).
  \bibitem{KURAEV} E.~A.~Kuraev and V.S.~Fadin,
    Yad.\ Fiz.\ \textbf{41}, 733 (1985) [Sov.\ J.\ Nucl.\ Phys.\ \textbf{41}, 466 (1985)].
    %\bibitem{RADCOR} F.~A.~Berends and R.~Kleiss, Nucl.\ Phys.\ B186, 22 (1981).
  \bibitem{BONNEAU} G.~Bonneau and F.~Martin,
    Nucl.\ Phys.\ B \textbf{27}, 381 (1971).
  \bibitem{GEANT4} S.~Agostinelli \emph{et al.},
    Nucl.\ Instrum.\ Methods Phys.\ Res., Sect.\ A \textbf{506}, 250 (2003).
  \bibitem{TRACKS} M.~N.~Achasov \emph{et al.} (SND Collaboration),
    JETP \textbf{101}, 1053 (2005).
  \bibitem{PDG} C.~Patrignani \emph{et al.} (Particle Data Group),
    Chin.\ Phys.\ C \textbf{40}, 10 (2016).
\end{thebibliography}
\end{document}